# HENNC: Hardware Engine for Artificial Neural Network-based Chaotic Oscillators

Mobin Vaziri[†], Shervin Vakili[‡], M. Mehdi Rahimifar[*], and J.M. Pierre Langlois[†]

[†] Department of Computer and Software Engineering, Polytechnique Montréal, Canada
[‡] Institut National de la Recherche Scientifique (INRS), Montréal, Canada
[*] Interdisciplinary Institute for Technological Innovation - 3IT, Université de Sherbrooke, Canada
mobin.vaziri@polymtl.ca

*Abstract*—This letter introduces a framework for the automatic generation of hardware cores for Artificial Neural Network (ANN)-based chaotic oscillators. The framework trains the model to approximate a chaotic system, then performs design space exploration yielding potential hardware architectures for its implementation. The framework then generates the corresponding synthesizable High-Level Synthesis code and a validation test-bench from a selected solution. The hardware design primarily targets FPGAs. The proposed framework offers a rapid hardware design process of candidate architectures superior to manually designed works in terms of hardware cost and throughput. The source code is available on GitHub[1].

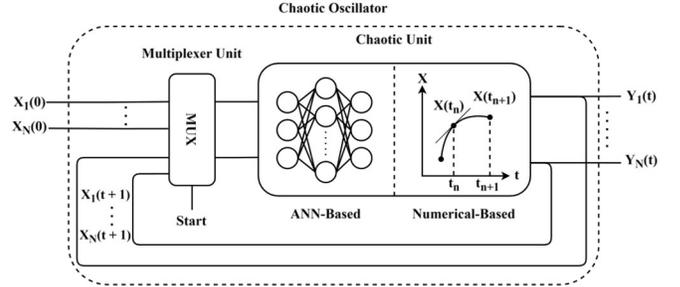

**Fig. 1.** Architecture of chaotic oscillators.

## I. INTRODUCTION

THIS letter is motivated by the need to counter cyber-attacks on image encryption algorithms [1]. For these applications, a high throughput Pseudo-Random Number Generator (PRNG) is required for real-time operation [2]. Chaotic oscillators are an appealing choice for PRNGs because of their complex dynamics, ergodicity, and sensitivity to initial conditions. Related studies have shown that, compared to traditional PRNGs like Linear Feedback Shift Registers (LFSRs), chaotic oscillators offer significantly higher resilience against attacks [3], [4]. Numerical methods such as Runge-Kutta (RK) can produce accurate solutions for even the most complex chaotic systems [5], but their hardware implementation is intricate and resource-intensive. Previous research has demonstrated the feasibility of using Artificial Neural Networks (ANNs) to approximate chaotic systems [6]. This letter thus introduces a framework to generate ANN-based chaotic oscillator architectures and corresponding hardware cores suitable for PRNG implementation. It also introduces novel latency and hardware cost estimation functions for fast design space exploration, allowing designers to produce High-Level Synthesis (HLS) model descriptions, suitable for Field Programmable Gate Arrays (FPGAs) implementation [7], in a matter of minutes.

## II. BACKGROUND

Fig. 1 illustrates a chaotic oscillator with a multiplexer (MUX) and a chaotic unit. Chaotic systems computations involve discretization through accurate but costly methods with a preference for Feedforward Neural Networks (FFNN) for chaotic system modeling [10]. However, these networks tend to accumulate feedback errors that affect the generated sequences. Al-Musawi et al.'s ANN model has fewer hidden neurons but high area consumption [11].

System performance hinges on the realization method within the chaotic system unit. The system parameters consist of the input and output neuron counts, contingent on the dimensionality of the chaotic system, along with the hidden neuron count, influencing the accuracy of the approximated curve. Zhang [6] also explored the optimal number of hidden neurons based on the Mean-Square Error (MSE). The findings revealed no improvement beyond eight hidden neurons. It also demonstrated the feasibility of modeling diverse chaotic systems using an ANN model. Yu et al. [12] provided evidence that ANN-based chaotic systems can produce random sequences that pass rigorous randomness tests in the NIST SP 800-22 test suite [13].

Consider a system of N differential equations:

$$\frac{dX_i}{dt} = f_i(X_1, X_2, \ldots, X_N), \qquad (1)$$

where $X_i$ are system variables, and t denotes time. The RK 4th-order (RK-4) method is calculated two steps: state variable coefficients (2) and variable update (3).

$$
\begin{aligned}
k_{1i} &= f_i(X_1, X_2, \ldots, X_N) \\
k_{2i} &= f_i\left(X_1 + \frac{dt}{2} \cdot k_{11}, \ldots, X_N + \frac{dt}{2} \cdot k_{1N}\right) \qquad (2) \\
k_{3i} &= f_i\left(X_1 + \frac{dt}{2} \cdot k_{21}, \ldots, X_N + \frac{dt}{2} \cdot k_{2N}\right)
\end{aligned}
$$

$$k_{4i} = f_i(X_1 + dt \cdot k_{31}, \ldots, X_N + dt \cdot k_{3N})$$

$$X_{i+1} = X_i + \frac{1}{6} \times (k_{1i} + 2 \times k_{2i} + 2 \times k_{3i} + k_{4i}) \quad (3)$$

The multiplications and additions for RK-4 in (2) and (3) can be divided into static and dynamic terms. Static terms encompass operations that calculate the input arguments in (2). Dynamic terms refer to operations present in the selected chaotic system, $f_i$. The total number of operations is:

$$
\begin{aligned}
n_{mul} &= n(mul_{static}) + n(mul_{dynamic}) \\
&= (3 \cdot N^2 + 3 \cdot N) + 4 \cdot n(mul_{dynamic}) \quad (4)
\end{aligned}
$$

$$
\begin{aligned}
n_{add} &= n(add_{static}) + n(add_{dynamic}) \\
&= (3 \cdot N^2 + 4 \cdot N) + 4 \cdot n(add_{dynamic}).
\end{aligned}
$$

The dynamic terms depend on the selected chaotic system. For instance, for the low-complexity Chen chaotic system [14]:

$$
\begin{cases}
\dfrac{dX_1}{dt} = a \cdot (X_2 - X_1) \\
\dfrac{dX_2}{dt} = (c - a) \cdot X_1 - (X_1 \cdot X_3) + c \cdot X_2 \\
\dfrac{dX_3}{dt} = X_1 \cdot X_2 - b \cdot X_3
\end{cases} \quad (5)
$$

there are six multipliers and five adders. On the other hand, ANN computation in the layer $l$ and $i^{th}$ neuron is given by:

$$Y_i^{(l)} = \phi \left( \sum_{j=1}^{n_{l-1}} W_{ij}^{(l)} X_j + b_i^{(l)} \right) \quad (6)$$

where $\phi$ is the activation function, $n$ is the number of neurons in the layer, and $w$ and $b$ are the weights and biases of the model, respectively. The number of operations in a general ANNs is obtained by:

$$n_{mul} = \sum_{i=2}^{L} n_i \times n_{i-1} \quad , n_{add} = \sum_{i=2}^{L} n_i \times (n_{i-1} + 1). \quad (7)$$

Table I compares the number of operations to implement a minimal chaotic system using the RK-4 method and an ANN proposed by Zhang [6]. ANNs execute a predetermined number of operations based on their structure. However, the computational complexity of the RK-4 method depends on the specific chaotic system it employs. These chaotic systems involve trigonometric or other non-linear functions, which are costly in resources. Hence, while achieving a negligible error (as presented in Table II) compared to the numerical methods, ANNs can offer stable and more efficient solutions than numerical methods.

TABLE I
Number of Operations in an ANN with 8 Neurons in the Hidden Layer, and in RK-4 with Chen Chaotic System

| Method | # Multiplications | # Additions |
|--------|------------------|-------------|
| ANN (3 – 8 – 3) [6] | 48 | 59 |
| RK-4 | 60 | 59 |

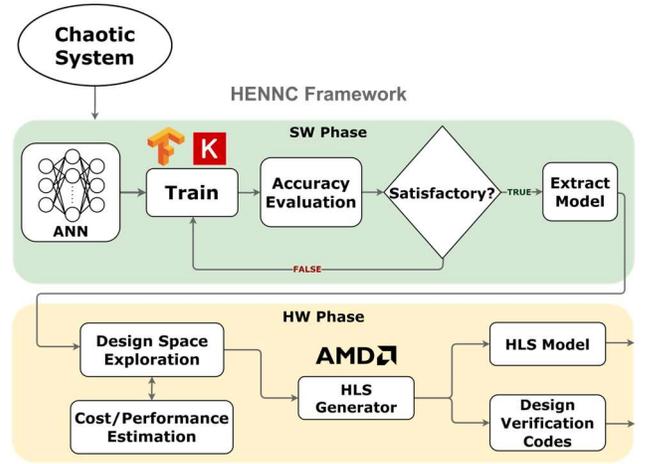

**Fig. 2.** HENNC framework design flow

## III. Proposed Framework

This section describes the two phases of the HENNC framework, illustrated in Fig. 2.

### A. Software Phase

The software phase of the HENNC framework commences with the generation of a chaotic sequences dataset, by numerically solving the defined chaotic system using the *Odeint* function from Python's *SciPy* library. We generate 100 k sequences and use 80% for training and 20% for test. To create the training dataset, we sample the output of the numeric chaotic system at each time step. Since the output at time step $t$ is the input to the system at time step $t$+1, each labeled data point consists of the output samples from two consecutive time steps. The dataset in the initial phase is used to train the ANN model, which is built with the *Keras* library, and approximates the chaotic system's function through regression. Users specify model hyperparameters, and the model's performance is evaluated with the Mean Absolute Error (MAE), Mean Square Error (MSE), Root Mean Square Error (RMSE), and R-squared ($R^2$). Table II presents a hyperparameters set for training the model for Chen's chaotic system. The training process terminates when the model achieves the desired accuracy, and the network parameters are extracted for the hardware phase.

### B. Hardware Phase

*1) Design Space Exploration:* Hardware design begins with the search for Pareto-optimal microarchitectures for the ANN model trained in the software phase. The HENNC framework incorporates a highly-configurable template HLS design with various model hyperparameters and a set of HLS directives that trade off parallelism and resource utilization. This flexibility allows the exploration of various ANN core microarchitectures.

HENNC offers users three options: minimum latency, lowest cost, and Pareto-optimal solutions with parallelism $P$. In the lowest cost solution ($P = 0$), the hardware core contains only one adder and one multiplier. Increasing parallelism reduces latency but increases costs. When $P \geq 1$, the number of multipliers and adders is ($2^P \times I$), with $I$ denoting the number of neurons in the input and output layers. For faster candidate solution characterization, HENNC incorporates a cost and throughput estimation method, described in the next subsection. Fig. 3a depicts the architectural design space, with the estimated cost and latency for a 3-16-3 ANN.

TABLE II
HYPERPARAMETERS AND PERFORMANCE OF ANN-BASED MODEL

| Hyperparameters | | |
|---|---|---|
| Loss Function | Optimizer | Learning Rate |
| MSE | Adam | $1 \times 10^{-4}$ |
| **Performance Metrics** | | |
| Activation Function | MSE | MAE | RMSE | $R^2$ |
| ReLU | 0.00031 | 0.01165 | 0.01755 | 0.99999 |
| Tanh | 0.00698 | 0.4434 | 0.08352 | 0.99982 |
| Sigmoid | 0.04412 | 0.10970 | 0.21004 | 0.99875 |

The hardware cores support single-precision floating-point computations and, by default, map these operations onto FPGA Digital Signal Processing (DSP) resources. Users have the option to utilize FPGA Look-Up Tables (LUTs) exclusively. Once the user selects the preferred candidate solution, the HENNC framework generates the corresponding chaotic oscillator HLS model.

*2) Estimation functions:* Pre-synthesis cost and speed estimation with HLS tools is time-consuming with limited accuracy. To tackle this issue, we developed cost and latency estimation models with an experimental approach. To estimate latency, we began by measuring the post-synthesis actual latency, in terms of the number of clock cycles, for a range of selected solutions across various ANN sizes and parallelism levels. After normalizing the latency results, we computed the average latency for each parallelism level. Using these averages, we calculated a polynomial interpolation of degree 3. The resulting latency estimation function is as follows:

$$Latency = (I \cdot H) \cdot (b_3 P^3 + b_2 P^2 + b_1 P + b_0), \quad (8)$$

where $b_3$ to $b_0$ are constant coefficients, and $P$ is the parallelism level. Due to substantial variations in latency with or without DSP utilization, the measurements and interpolation calculations described above are carried out separately for the two scenarios, yielding distinct value sets for $b_3$ to $b_0$. Fig. 3b presents the actual latency results when utilizing DSP resources, shown in terms of the number of clock cycles normalized by dividing them by $I \times H$. The MATLAB Curve Fitter Toolbox was used to determine the interpolation function and the coefficients $b_3$ to $b_0$. This process was repeated separately for LUT-based solutions.

For the cost estimation, we generated and synthesized a range of solutions to gain insights into how LUT utilization varies with hyperparameter values and parallelism. LUT usage was measured as we incremented the number $H$ of hidden layer neurons and the number $I$ of neurons in the input/output layers, while using two levels of parallelism. There is a semi-linear relationship between LUT utilization and both $H$ and $I$. However, increasing the level of parallelism exerts a nonlinear influence on the hardware core control circuit. we opted to express the estimation of LUT usage as a linear function of $I$ and $H$ for each parallelism level ($P$), employing the following formulation:

$$\#LUT = (c_1 \cdot I \cdot H) + (c_2 \cdot I) + (c_3 \cdot H) + \beta, \quad (9)$$

where $c1$, $c2$, $c3$, and $\beta$ are coefficients determined experimentally for each parallelism level.

These coefficients were established through experimentation for each unique parallelism level, with a curve fitting tool. For every potential parallelism level, we conducted linear interpolation of LUT vs. $H$ and LUT vs. $I$ results from actual result curves, such as those depicted in Fig. 4. Subsequently, we determined the coefficient values by equating Eq. 9 with the linear interpolation results. The obtained coefficient values for various parallelism levels were then compiled to create a constant coefficient table. When assessing the LUT cost for a specific candidate architecture, the HENNC framework retrieves the $c_1$, $c_2$, $c_3$, and $\beta$ values associated with the requested parallelism level from this table. These values are then incorporated into Eq. 9 to derive the estimation function with two variables, $I$ and $H$.

*3) Hardware core generation:* In the final step, the HENNC framework generates the HLS design and verification codes for the user-selected solution. Two distinct C++ code files are generated: one for the synthesizable hardware design and another for the testbench to aid in design validation through co-simulation in AMD-Vitis HLS.

## IV. RESULTS

We used AMD-Xilinx Vitis for HLS synthesis, AMD-Xilinx Vivado ML 2023.1 for RTL synthesis, targeting a xcku035 Kintex Ultrascale FPGA with a speed grade of -3. Table III lists the hardware costs for the three most commonly used ANN models for approximating 3-D chaotic systems. It shows that all HENNC solutions require significantly fewer LUTs and operate at considerably higher clock frequencies compared to [9], [11]. In contrast to the architecture proposed by Alcin et al. [9], HENNC's fastest architecture offers reduced latency while consuming nearly 20× fewer LUTs. The other two HENNC solutions have increased latency but lower hardware costs. A primary reason for the substantial decrease in LUTs is the

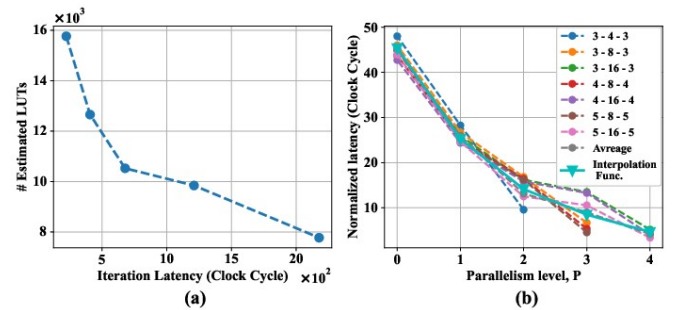

**Fig. 3.** (a) Estimated cost and latency for the 3-16-3 ANN. (b) Normalized actual latencies and the interpolation curve when utilizing DSP resources.

## TABLE III
### Hardware Costs for HENNC Candidate Designs and Existing Works.

| ANN | P | Solution | HENNC LUTs With DSP | HENNC LUTs No DSP | HENNC DSPs With DSP | HENNC Iter Lat (CC) With DSP | HENNC Iter Lat (CC) No DSP | Post LUTs With DSP | Post LUTs No DSP | Post FFs With DSP | Post FFs No DSP | Post DSPs With DSP | fmax With DSP | fmax No DSP | Post Iter Lat (ns) With DSP | Post Iter Lat (ns) No DSP |
|---|---|---|---|---|---|---|---|---|---|---|---|---|---|---|---|---|
| 3 − 4 − 3 | 2 | 1 | 4100 | 10708 | 44 | 169 | 128 | 4255 | 11358 | 8754 | 14909 | 44 | 510.46 | 548.2 | 224.3 | 167.4 |
|  | 1 | 2 | 2288 | 5592 | 22 | 302 | 253 | 3683 | 7089 | 7383 | 10334 | 22 | 510.46 | 548.2 | 661.1 | 516.9 |
|  | 0 | 0 | 2215 | 2896 | 5 | 544 | 500 | 2853 | 3314 | 5635 | 5735 | 5 | 510.46 | 548.2 | 1123.2 | 953.7 |
| 3 − 8 − 3 | 3 | 1 | 7988 | 21204 | 88 | 203 | 165 | 8051 | 22310 | 15020 | 27439 | 88 | 510.46 | 548.24 | 312.0 | 227.5 |
|  | 2 | 2 | 5744 | 12352 | 44 | 339 | 256 | 5885 | 12279 | 9981 | 14210 | 44 | 494.55 | 387.14 | 816.1 | 717.2 |
|  | 1 | 3 | 5180 | 8484 | 22 | 605 | 506 | 5089 | 8307 | 7944 | 10001 | 22 | 494.55 | 387.14 | 1087.8 | 1017.7 |
|  | 0 | 4 | 4067 | 4721 | 5 | 1089 | 1001 | 3962 | 4440 | 6388 | 6485 | 5 | 494.55 | 386.10 | 1848.1 | 1880.3 |
| 3 − 16 − 3 | 4 | 1 | 15764 | 42196 | 176 | 223 | 290 | 15671 | 43604 | 28723 | 53538 | 176 | 510.46 | 548.24 | 483.6 | 344.0 |
|  | 3 | 2 | 12656 | 25872 | 88 | 407 | 330 | 11868 | 24181 | 19592 | 27516 | 88 | 494.55 | 370.23 | 1309.0 | 1144.8 |
|  | 2 | 3 | 10524 | 17132 | 44 | 678 | 513 | 10271 | 16511 | 14688 | 18338 | 44 | 494.55 | 387.14 | 1571.6 | 1377.7 |
|  | 1 | 4 | 9841 | 13145 | 22 | 1211 | 1013 | 9846 | 12844 | 16818 | 19396 | 22 | 510.46 | 548.24 | 2427.8 | 1914.6 |
|  | 0 | 5 | 7771 | 8425 | 5 | 2178 | 2003 | 7726 | 7795 | 15051 | 14759 | 5 | 510.46 | 548.24 | 4212.1 | 3618.2 |
| 3 − 4 − 3 [11] | - | 1 | - | - | - | - | - | 21172 | - | 96 | - | 162 | 7.76 | - | - | - |
| 3 − 8 − 3 [9] | - | 1 | - | - | - | - | - | 87207 | - | 86329 | - | 8 | 266.43 | - | 543.8 | - |

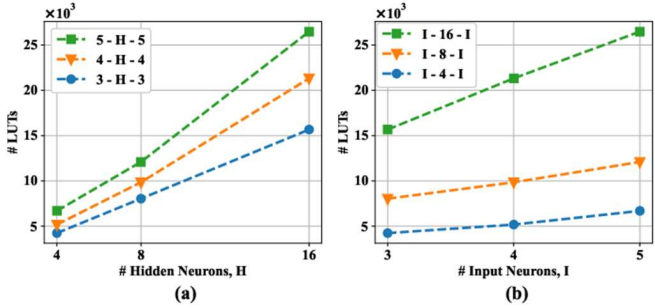

**Fig. 4.** Post-synthesis number of LUTs as a function of the number of neurons in the (a) hidden layer, (b) input and output layers.

adoption of ReLU as an activation function. In contrast, prior works such as [11] employed the exponential function, and [9] utilized numerous floating-point dividers and CORDIC cores to approximate the activation function. For a 3-8-3 ANN, an Intel 12th Gen Intel(R) Core(TM) i7-12700 CPU generates 100 output samples in approximately 3.5 seconds, while the FPGA-based design accomplishes the same task in 13 microseconds.

Fig. 5 represents the HENNC design space for different ANN sizes with and without DSPs. The figure illustrates the trade-offs between post-synthesis LUT utilization and the latency offered by each candidate solution. For each ANN size, the smallest and largest solutions in terms of LUT usage represent the top-speed and cost-optimized solutions, respectively. The top-speed hardware leverages maximum parallelism by employing the maximum number of MAC operators. In contrast, the cost-optimized mode typically deploys only a single MAC unit to calculate all neurons.

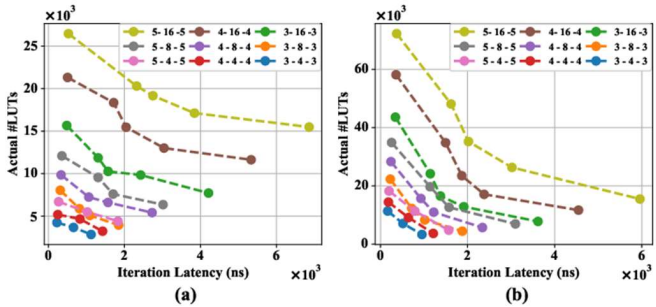

**Fig. 5.** Post-synthesis hardware cost and latency of the design space explored by HENNC for different ANN sizes in two modes: (a) with DSP utilization, and (b) without DSP utilization.

## References

[1] Z. Hua, Y. Zhou, and H. Huang, "Cosine-transform-based chaotic system for image encryption," *Information Sciences*, 2019.

[2] R. Hamza, Z. Yan, K. Muhammad, P. Bellavista, and F. Titouna, "A privacy-preserving cryptosystem for iot e-healthcare," *Information Sciences*, vol. 527, pp. 493–510, 2020.

[3] S. Kalanadhabhatta, D. Kumar, K. K. Anumandla, S. A. Reddy, and A. Acharyya, "Puf-based secure chaotic random number generator design methodology," *IEEE transactions on very large scale integration (VLSI) systems*, vol. 28, no. 7, pp. 1740–1744, 2020.

[4] M. N. Aslam, A. Belazi, S. Kharbech, M. Talha, and W. Xiang, "Fourth order mca and chaos-based image encryption scheme," *IEEE Access*, vol. 7, pp. 66 395–66 409, 2019.

[5] K. Rajagopal, M. Tuna, A. Karthikeyan, 'I. Koyuncu, P. Duraisamy, and A. Akgul, "Dynamical analysis, sliding mode synchronization of a fractional-order memristor hopfield neural network with parameter uncertainties and its non-fractional-order fpga implementation," *European Physical Journal Special Topics*, vol. 228, pp. 2065–2080, 2019.

[6] L. Zhang, "Artificial neural network model design and topology analysis for fpga implementation of lorenz chaotic generator," in *Canadian Conf. on Electrical and Computer Engineering*, 2017, pp. 1–4.

[7] A. HajiRassouliha, A. J. Taberner, M. P. Nash, and P. M. Nielsen, "Suitability of recent hardware accelerators (dsps, fpgas, and gpus) for computer vision and image processing algorithms," *Signal Processing: Image Communication*, vol. 68, pp. 101–119, 2018.

[8] M. Alcin, "The runge kutta-4 based 4d hyperchaotic system design for secure communication applications," *Chaos Theory and Applications*, vol. 2, no. 1, pp. 23–30, 2020.

[9] M. Alc¸ın, 'I. Pehlivan, and 'I. Koyuncu, "Hardware design and implementation of a novel ann-based chaotic generator in fpga" *Optik*, 2016.

[10] J. Sunny, J. Schmitz, and L. Zhang, "Artificial neural network modelling of rossler's and chua's chaotic systems," in *IEEE Canadian Conference on Electrical & Computer Engineering*, 2018, pp. 1–4.

[11] W. A. Al-Musawi, W. A. Wali, and M. A. A. Al-Ibadi, "New artificial neural network design for chua chaotic system prediction using fpga

hardware co-simulation," *International Journal of Electrical and Computer Engineering*, vol. 12, no. 2, p. 1955, 2022.

[12] F. Yu, Z. Zhang, H. Shen, Y. Huang, S. Cai, J. Jin, and S. Du, "Design and fpga implementation of a pseudo-random number generator based on a hopfield neural network under electromagnetic radiation," *Frontiers in Physics*, vol. 9, 2021.

[13] A. Rukhin, J. Soto, J. Nechvatal, M. Smid, E. Barker, S. Leigh, M. Levenson, M. Vangel, D. Banks, A. Heckert et al., A statistical test suite for random and pseudorandom number generators for cryptographic applications. *US Department of Commerce, Technology Administration*, 2001, vol. 22.

[14] M. S. Azzaz, C. Tanougast, S. Sadoudi, R. Fellah, and A. Dandache, "A new auto-switched chaotic system and its fpga implementation," *Communications in Nonlinear Science and Numerical Simulation*, 2013.